# Influence of Mg doping on In adsorption and In incorporation in (In,Ga)N superlattices

Erdi Kuşdemir (1), Caroline Chèze(1)[*], and Raffaella Calarco (1 and 2)

( (1) Paul-Drude-Institut für Festkörperelektronik, Leibniz-Institut im Forschungsverbund Berlin e.V., Berlin, Germany, (2) Consiglio nazionale delle ricerche - CNR Institute for Microelectronics and Microsystems -IMM, Rome, Italy)

We present a detailed investigation of the mechanisms at play for the incorporation of In and Mg on the GaN(0001) surface during plasma-assisted molecular beam epitaxy (PAMBE). First, we have studied the kinetics of In desorption in the presence of Mg either without or with N supply from the plasma cell by quadrupole mass spectrometry (QMS) in the line of sight. Second, we have explored the effect of Mg doping at a different time along the cycle of (InN/GaN) supply repeated 10 times to form (In,Ga)N/GaN superlattices (SLs). By the complementary ex-situ investigation of these SLs by X-ray diffraction (XRD) and secondary ion mass spectrometry (SIMS), we found that in monolayer-thick (In,Ga)N layer, the In content was maximized when Mg was not supplied simultaneously to In, but it drastically decreased otherwise. In contrast, the Mg concentration strongly increased in the (In,Ga)N monolayers compared to the GaN barriers. We attribute this finding to the surfactant effect of In for Mg, which decreases the binding energy of Mg in GaN in presence of N.

* cheze@pdi-berlin.de

I. INTRODUCTION

As-grown GaN always displays n-type conductivity. To achieve efficient p-type doping and realize a p-n junction and eventually a light-emitting diode (LED) or a laser diode (LD), unintentional n-type conductivity has to be overcome by the hole concentration.[1] However, the large lattice mismatch between GaN and host foreign substrate yields high dislocation density that not only acts as non-radiative recombination centers but also may be the source for high unintentional n-type concentration.[2] Additionally, due to the high binding energy of Mg (~200 meV for GaN), only about 1% of Mg atoms are ionized at room temperature (RT) and free a hole. Therefore, to obtain a hole concentration of $10^{18}$ cm$^{-3}$, a Mg concentration of $10^{20}$ cm$^{-3}$ is required.[2] However, above the Mg concentration of $10^{20}$ cm$^{-3}$, the Fermi level stabilizes at a level that is more favorable to the formation of donor-like intrinsic defects that yield compensation.[2,3] Thus, Mg-doping in GaN(0001) crystal is commonly carried out in a narrow doping window, which limits the ease of III-Nitride applications.

To widen the doping window of GaN restricted by donor-like defects, metal modulated epitaxy, that interchanges the Mg and Ga supplies, was proposed as an alternative method, as it allowed to push the Mg concentration above $7 \times 10^{20}$ cm$^{-3}$.[4,5] Another approach used to increase the hole concentration is to grow Mg-doped (In,Ga)N instead of Mg-doped GaN, as the Mg activation energy decreases with the In mole fraction.[6] This approach presents the additional advantage that the Mg sticking coefficient increases by decreasing the growth temperature that is necessary to form high quality (In,Ga)N:Mg layers under In-rich regimes.[7] However, (In,Ga)N growth still presents several challenges. The alloy disorder in the ternary (In,Ga)N random alloy generally impedes the mobility of charge carriers due to scattering.[8] This is particularly an issue for the hole transport which limits the number of quantum wells (QWs) in the active region of LEDs or LDs.[9] In addition, it is crucial to avoid the accumulation of Mg and In at the growth surface.[10,11]

By the modulation of In, Ga and Mg supplies under constant N supply, control on the arrival rate of these species can be achieved. Moreover, with this technique superlattices (SLs) can be formed. When their period is as short as few monolayers, these SLs can even constitute an ordered alloy along the growth direction as they consist of stacked integer layers of (In,Ga)N and GaN. Therefore, these short-period SLs (SPSLs) should allow eliminating the issues related to thick (In,Ga)N layers, such as composition pulling effect and phase separation.[12] Furthermore, (In,Ga)N SLs doped with Mg may exhibit an increased average hole concentration compared to thick (In,Ga)N layers doped with Mg owing to the periodic modulation of the valence band edge, as it was demonstrated for (Al,Ga)N/GaN SLs.[13] Such heterostructures could be also implemented as a hole injection layer (HIL) or a hole accumulation layer (HAL) and help to overcome issues related to restricted hole injection, carrier delocalization, and electron leakage current.[14–16]

Yet, although the fabrication of InN/GaN SPSLs was demonstrated by Yoshikawa *et al.*[17], more recent studies have concluded that (In,Ga)N/GaN SPSLs formed, in which the QWs have a self-limited thickness of one monolayer and a maximum In content of 0.3 instead of 1.[18–21] To date, the fabrication and the properties of (In,Ga)N/GaN SLs doped with Mg have not been studied. In the literature, some work on p-type conductivity of thick InN:Mg and (In,Ga)N:Mg layers has been reported,[22–26] but little is known concerning the effect of Mg on the In incorporation in (In,Ga)N:Mg and on (In,Ga)N / GaN:Mg SLs. Investigations of thick layers of (In,Ga)N doped with Mg[6,27] have revealed that high Mg surface densities can disrupt the In incorporation. Since both Mg and In occupy the same cation sites in GaN matrix, the formation of a single monolayer of (In,Ga)N might also be strongly affected by the presence of Mg.

In the present study, we get a deeper understanding of the intricate kinetics of Mg and In incorporation at the GaN(0001) surface in order to form (In,Ga)N/GaN SLs with different modulation-doping schemes of Mg. First, we have monitored *in situ* the In desorption during the supply of different sequences of In, Ga, and Mg without and with N on *c*-plane (0001) GaN by quadrupole mass spectrometry (QMS) in the line of sight in a plasma-assisted molecular beam epitaxy (PAMBE) system. Second, we have investigated (In,Ga)N / GaN SLs doped with Mg by X-ray diffraction (XRD) and secondary ion mass spectrometry (SIMS). These SLs were formed by supplying (InN/GaN) periods with an Mg supply at a different time of the period.

## II.  EXPERIMENTAL

We have conducted principally two sets of experiments in a PAMBE system DCA P600 equipped with standard effusion cells and a radio frequency plasma source (Addon) for the supply of group-III elements and active nitrogen, respectively. The first set was devoted to investigating *in-situ* the effect of Mg on the In adsorption in vacuum or in the presence of N by QMS. The second set of experiments has been designed to yield the fabrication of structures to be externally characterized by XRD and SIMS.

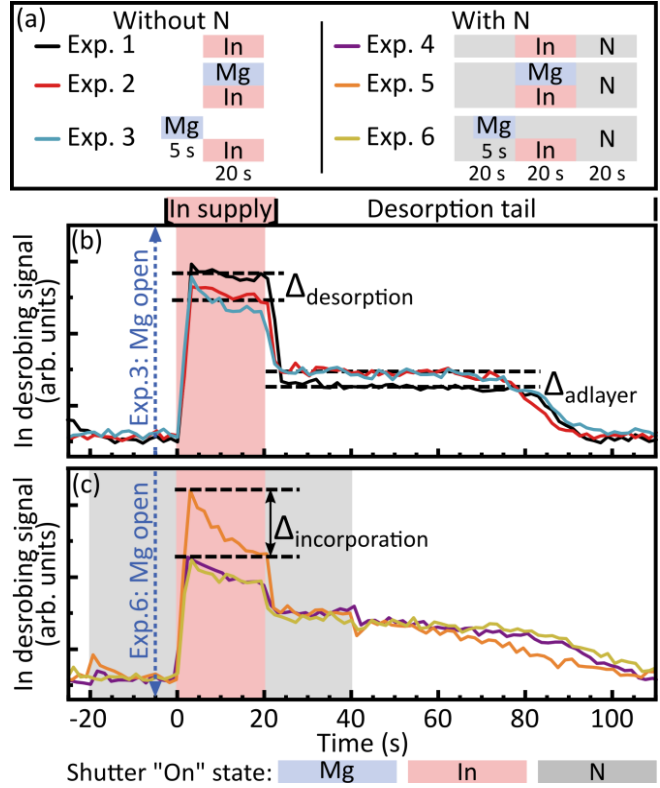

**Figure 1:** (a) Flux supply sequences for the experiments 1 – 6. (b) and (c) $^{115}$In desorption signal monitored with different flux supply sequences (b) in vacuum, and (c) in presence of N. In (c) a small increase (decrease) of the $^{115}$In signal is observed just after the N shutter opened (closed). This is an artifact stemming from the correction for the offset in the In desorption rate as the N plasma opened and closed. In (a) − (c) the supply sequences for the N, In, and Mg fluxes monitored by QMS are represented by gray, pink and blue bars. The red shaded area corresponds to the In supply time. The gray shaded area indicates the N supply time. In (b), the difference between the In desorption levels for the In supply and the desorption tail is labeled with "$\Delta_{desorption}$" and "$\Delta_{adlayer}$" and in (c) the difference between the In desorption levels is labeled "$\Delta_{incorporation}$".

In the first set of *in-situ* experiments, the Mg, In, and N fluxes were supplied in different orders on a GaN(0001) template while line-of-sight QMS[19] was utilized to measure the desorption signal of $^{69}$Ga, $^{115}$In, $^{24}$Mg, $^{25}$Mg, and $^{26}$Mg from the substrate surface. However, for the $^{69}$Ga, $^{24}$Mg, $^{25}$Mg, and $^{26}$Mg signals monitored by QMS, no change was observed due to the relatively low growth temperature, and the shadowing effect by the $^{28}$N$_2$ signal for the Mg species. Consequently, only the $^{115}$In signal is reported below. The Mg flux was set to 4.2 × 10$^{-8}$ Torr, and the N and In supply rates set to 9.0 nm/min and 7.3 nm/min in GaN growth rate units, respectively, with 1 ML corresponding to 1.136×10$^{15}$ atoms/cm$^2$ [GaN(0001) areal density]. The substrate temperature was set at 580 °C. The N shutter was opened 20 s before and closed 20 s after the metal flushes in order to stabilize the QMS signal. Indeed, opening the N shutter induced a change in the partial pressure and it yielded an offset in the record of the desorption signal of all species that were supplied or not. Opening the N shutter 20 s prior to opening the shutters of the metal cells allowed to disentangle this artifact from the measurement. The offset was corrected after acquisition in the curves presented in Figure 1. The In shutter was open for 20 s (12 MLs). The nominal ML thicknesses were calculated from the growth rates determined from thick layers grown under both metal- and N-limited regimes. Additionally, we used the parameters of wurtzite InN (and GaN), $c/2 = 0.288$ nm (0.259 nm) and refer to 1 ML along [0001].

The second set of experiments consisted of (In,Ga)N/GaN SLs formed with 10 repeats of (20 MLs InN/ 40 MLs GaN) sequences supplied at 610 °C following the approach described in ref. [19], where Mg was inserted at a different time in the sequence for each sample. Note that former studies have shown that the deposition of thick binary InN QWs at such high growth temperature yield actually QWs with a self-limited thickness of 1 ML of (In,Ga)N and a maximum In content of 0.33[18–21]. For these experiments, the N, In, and Ga growth rates were fixed to 10.0 nm/min, 8.7 nm/min, and 10.8 nm/min in GaN growth rate units and the Mg flux was set to $6.2 \times 10^{-9}$ Torr that corresponds to a nominal concentration of $8 \times 10^{20}$ cm$^{-3}$ML$^{-1}$ according to prior SIMS measurement on thick (In,Ga)N layers grown at the same temperature. All the samples were grown on commercially available Al$_2$O$_3$/n-type GaN(0001) templates with dislocation density lower than $5 \times 10^8$ cm$^{-2}$. Prior to the SLs, a GaN layer with a thickness of 100 nm was grown under intermediate Ga-rich conditions at 690°C.

The XRD data were acquired by PANalytical X'Pert system with CuKα[1] radiation using a Ge(220) hybrid monochromator, and a Ge(220) analyzer crystal for diffraction measurements. In order to get a deeper understanding of the effect of Mg supply on the structural properties of the samples under investigation, PANalytical X'Pert Epitaxy software was used to simulate the ω-2θ scans. Satisfactory fits were achieved by substituting the layers doped with Mg by virtual (Al,In,Ga)N layers with Al content 0.55 which was determined from the comparison of the fits of the undoped sample A and the one of an additional sample of GaN:Mg grown at the same temperature and with the same fluxes (not shown). The In content was then kept as a variable. SIMS measurements were performed at RTG Mikroanalyse GmbH to verify the Mg concentration and the In content in the (In,Ga)N/GaN SLs with ± 1 nm accuracy in depth. Note that the first 20 nm of the measurements were not reliable due to the stabilization of the instrument and surface charging effect.

### III. Investigation of Mg effect on In desorption without and with Nitrogen supply

Figure 1(a) sketches the different sequences of the In, Mg and N fluxes supplied during the six experiments of the first set. For experiments 1 – 3, either In, or In and Mg fluxes were supplied in vacuum, while experiments 4 – 6 reproduced experiments 1 – 3 with an additional supply of N. Experiment 1 which corresponds to the supply of only In on the GaN(0001) surface was utilized as a control experiment. Experiment 2 was designed to investigate the effect of Mg on In adsorption when both metal species are supplied simultaneously. Finally, in experiment 3, the GaN surface was first exposed to the Mg flux for 5 s, and after closing the Mg shutter, subsequently to the In flux. Figure 1(b) presents the $^{115}$In desorption signal acquired during experiments 1 – 3 in vacuum, while Figure 1(c) presents the results for experiments 4 – 6 under N supply. The desorption lines were superimposed by taking the opening of In shutter as time origin. For all experiments, we distinguished two different stages described as "In supply" and "desorption tail" that are highlighted in Figure 1(b) and (c). During the In supply stage, as the In shutter opened (time 0 s), a sudden increase of the

$^{115}$In signal to a maximum level was observed. For experiment 1, the maximum level was maintained as long as In was supplied. After closing the In shutter, the "desorption tail" stage set in with the immediate decrease of the In signal to an intermediate level that was attained in 5 s. The In signal returned to the background level after about 75 s. This measurement is explained as follows. For $^{115}$In, it has been already reported that the maximum desorption level measured by QMS is dependent on the substrate temperature and follows an Arrhenius law.[28] At the temperature of 580 °C, the supplied In rate exceeds the maximum desorption rate possible (0.9 nm/min),[28] and In atoms that cannot desorb, accumulate at the surface in the form of an adlayer. After the In shutter is closed, the desorption of the adlayer occurs during the desorption tail stage extending until the signal recovers its background level. In the case of experiments 2, and 3, no significant difference to experiment 1 was observed. The In signal reached the same value in experiments 1 – 3 as the In shutter was opened. However, in contrast to experiment 1, for experiments 2 and 3, the In signal slightly decreased during the In supply stage which is highlighted as "$\Delta_{desorption}$" in Figure 1(b). Therefore, the presence of Mg might have increased the coverage of In on GaN(0001) surface by increasing the probability of finding an adsorption site.[29] After the In shutter was closed, during the desorption tail stage, there was a slight increase in the desorption level of the accumulated In adlayer ($\Delta_{adlayer}$). Note that the slight decrease of the In supply level and the slight increase of the desorption tail level for experiments 2 and 3 compared to experiment 1 is inside the error margin of the measurement. However, the calculation of $\Delta_{desorption}$ and $\Delta_{adlayer}$ coverage by integration below the curve for individual repetition led to similar results, which supports the idea that the In absorption was increased by the presence of Mg.

In experiments 4 – 6, where N was additionally supplied, the two stages of the In desorption signal described above are also presented in Figure 1(c). Two significant differences were observed, that contrast with the desorption experiments in vacuum. The first change occurred during the "supply" stage, where the maximum level of the In desorption signal in experiment 5 (In and Mg supplied simultaneously) was increased by 15% compared to the level measured in experiments 4 (without Mg supply) and 6 (Mg and N supplied prior to In). This finding possibly stems from the inhibition of In adsorption in presence of Mg and N in the competition between In and Mg for the cation adsorption sites at N dangling bonds.[26,30] Note that a similar effect was first reported for the growth of GaN:Mg.[3] In contrast, for experiment 6, Mg was supplied before In and it possibly fully incorporated with N into the GaN matrix when the In shutter opened. Thus, during the subsequent supply of In, the In desorption was not affected by the presence of Mg and attained the same level as in experiment 4 where no Mg was supplied.

The second important difference with the desorption experiments 1 – 3 in vacuum was the identity of the In desorption level observed for experiments 4 – 6 during the "desorption tail" stage. This finding may be related to the fact that Mg was supplied in the presence of N and it fully incorporated in the crystal and was not present in the adlayer. Thus, the level shift observed for the desorption tail of experiments 1 – 3 was not observed for experiments 4 – 6.

## IV. Investigation of Mg and In incorporation in Mg-doped (InGaN/GaN) SLs

### A. *In-situ* monitoring of the In and Mg desorption by QMS

In order to correlate the impact of Mg on the In incorporation with the In desorption mechanisms observed *in situ*, a set of SL samples (samples A – E) was grown and characterized *ex-situ*. Figure 2(a) presents the nominal structure of these samples. Sample A was a non-intentionally doped SL. In contrast, samples B – E had the same nominal structure as sample A, but Mg was supplied either during the whole period (sample B), the GaN barrier only (sample C), the first 20% of the GaN barrier (sample D), or the QW only (sample E). The In desorption signal was also recorded *in situ* by line-of-sight QMS during the whole growth time. The results of the measurement are shown in Figure 2(b) – 2(d). For all these samples, again the two stages of the In desorption described above were observed, i.e. the In desorption signal reached a maximum level, and stayed steady as long as the In shutter remained open (In supply). Then, upon closing the In cell, the In level dropped with time and returned to the background level (desorption tail).

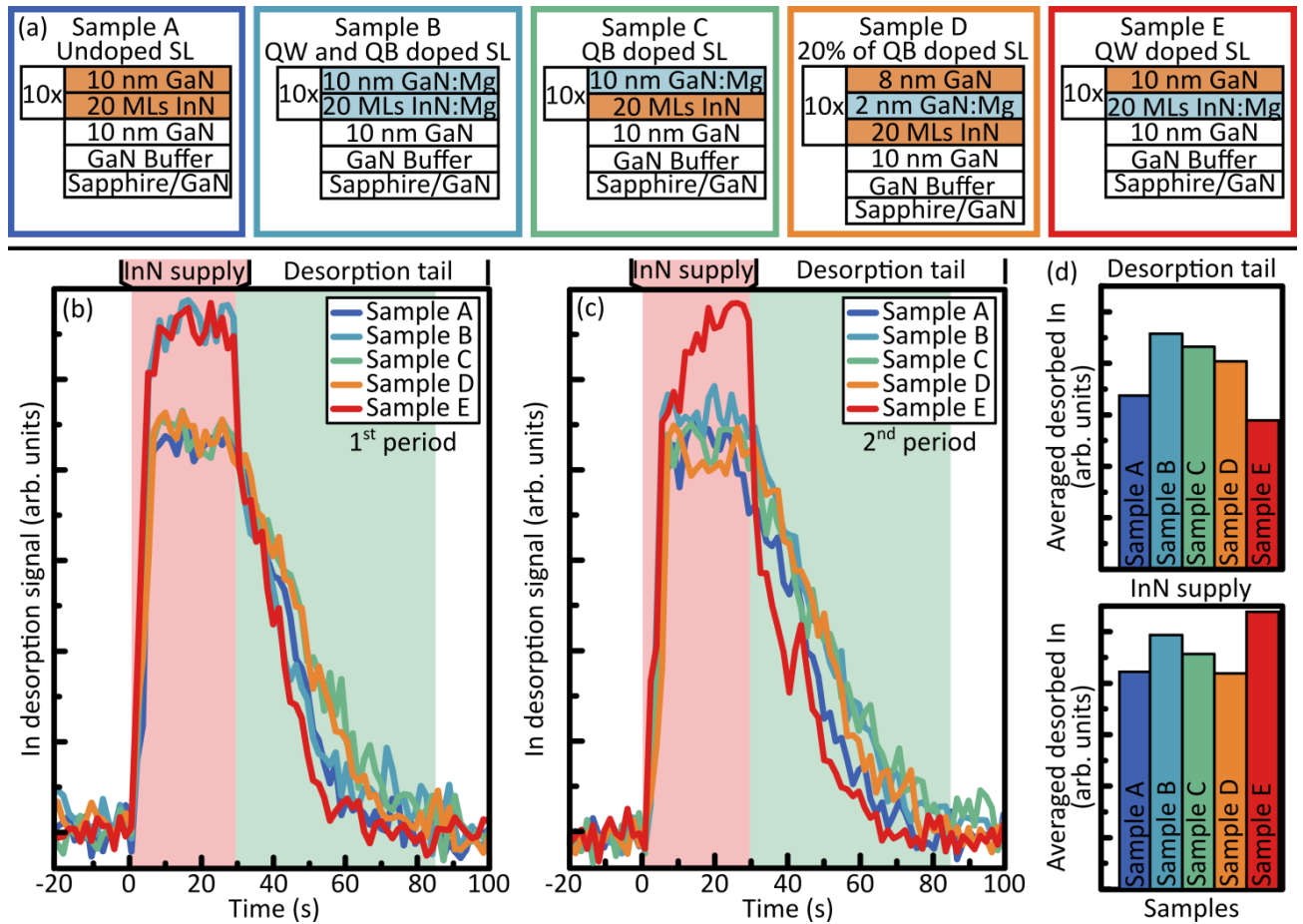

**Figure 2:** (a) Sketch of the nominal sample structures of the second set. The samples A – E are SLs with ten periods as described in the text. The undoped layers and the Mg-doped layers are represented by the orange and blue shaded areas, respectively. $^{115}$In desorption signal of the (b) first, and (c) second periods of the SL samples A – E monitored by line-of-sight QMS. The open state of the In and the Ga shutters are marked with a red and green shaded area, respectively. (d) Amount of In desorbed during the InN supply stage and the desorption tail stage averaged over $2^{nd}$ – $10^{th}$ periods.

Figure 2(b) presents the $^{115}$In desorption signal over the first period of growth. Similar to experiments 4 – 6 of the first set, for the samples of the second set the In desorption reached the same maximum level as the undoped sample (sample A) except for samples B and E, which are those for which Mg was supplied simultaneously to In. Compared to sample A, the In desorption maximum for sample B and E was increased by 27% and 28%, respectively. Again, this result may be attributed to the fact that both Mg and In occupy the same cation sites in GaN.[2,3,31] When Mg and In are supplied simultaneously, Mg may segregate at the surface and inhibit the occupancy of In at the surface, similar to the Mg surfactant effect reported during the growth of GaN(0001) doped with Mg.[32] In turn, the In incorporation is reduced and the In desorption increases as already observed for experiment 4. However, note that for the $2^{nd} – 10^{th}$ periods [Figure 2(c)] the maximum In desorption level during the InN supply stage for sample B, dropped to a level similar to the one of sample A. A possible explanation is that during the GaN:Mg growth of previous periods, unincorporated Mg atoms may have segregated and accumulated on the surface. Therefore, during the subsequent In supply, the segregated and supplied Mg atoms were most probably present in the adlayer and increased the In adlayer coverage as observed for experiments 2 and 3 of the first set. Furthermore, this increase in In coverage may have yield the decrease of the In desorption to level similar to the one of sample A. During the desorption tail stage, the In adlayer desorbed with a maximum level comparable for all samples [Figure 2(d)].

Figure 2(d) presents the In desorption averaged over the $2^{nd}$ to the $10^{th}$ periods. The area below the In desorption signal recorded by QMS, was integrated on the one hand over the time of the InN supply, and on the other hand over the time for which the desorption tail was observed. These results confirmed the previous observations. During the InN supply stage, the amount of In desorbed for the samples where In and Mg were supplied simultaneously (samples B and E) increased by 15% and 23% compared to the undoped sample (sample A) while a similar amount to samples A was measured for the rest of the samples. As mentioned before, this result is possibly attributed to the fact that both Mg and In occupy the same cation lattice sites in GaN.[2,3,31] Furthermore, over the desorption tail stage, the amount of In desorbed of samples B, C, and D increased by 26%, 22%, and 16% compared to the one of sample A, respectively. As for all these samples, Mg was supplied during the desorption tail stage, the increased In desorption is possibly related to the accumulation and segregation of Mg at the GaN surface. In contrast, for sample E the amount of In desorbed during the tail stage decreased by 11% relative to sample A possibly because the In adlayer coverage was already decreased due to the presence of Mg during the In supply stage. In order to verify the effect of Mg on In desorption measured by QMS, *ex-situ* XRD and SIMS characterization was conducted.

**B.   Structural characterization by XRD**

Figure 3 displays the $\omega$-$2\theta$ scans around the GaN(0002) reflection of the samples A – E. All the samples presented a sharp $Al_2O_3$(0006) peak at 20.85° and the GaN(0002) peak at 17.28° related to the $GaN/Al_2O_3$ templates, as well as $SL_n$ satellite peaks on both sides of the GaN(0002) reflection characteristic for the structure periodicity. Though the periodicity of the satellites was mainly determined by the GaN barrier thickness, the number, width, and intensity of the satellites varied from sample to sample, which is to relate to different In incorporation and interface roughness of the SL structures. For the samples where Mg was supplied simultaneously to InN (samples B and E), the intensity of the positive $SL_{+n}$ satellite set was higher than the intensity of the $SL_{-n}$ satellite set at an angle lower than $SL_0$. In the analytical approach, the direct relationship between the intensity of the SL peak and the atomic plane spacing can be derived from equation 5 given in ref. [33]. Thus, the relative intensity of the $SL_{+n}$ set compared to the one of the $SL_{-n}$ set gives the information on the local atomic plane spacing. This finding possibly indicates a compressive strain effect of the Mg incorporation in these samples. In contrast, for the samples where Mg was supplied simultaneously to GaN (samples C, and D), the relative intensity of the $SL_{-n}/SL_{+n}$ set is the same as for the undoped sample A, i.e. the strain status should be similar. Besides, samples B and D had broader satellite peaks with less intensity. This result indicates the increased interface roughness of these heterostructures.

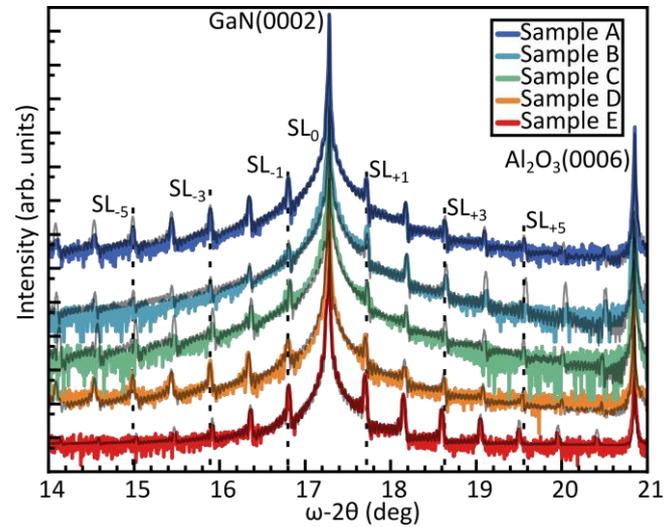

**Figure 3:** ω - 2θ scans of samples A – E around the GaN(0002) reflection. The scans have been shifted vertically for clarity. The simulation curves (gray) are superimposed with the measurement data in colors. $SL_n$ labels refer to the order of the satellite peaks. The $Al_2O_3$(0006) and GaN(0002) peaks are also labeled in the graph.

Dynamical simulations were done utilizing the X-Pert software to assess the In content assuming fully strained structures. This point was also verified by reciprocal space maps of the samples (not shown here). Table I presents the results of the simulations. As it was not possible to simulate directly (In,Ga)N:Mg layers, fictive (Al,In,Ga)N layers with fixed Al and varying In contents and with the thickness of 1 ML were utilized in the model, which provided satisfactory fit of the spectra of samples B and E. This method brought a qualitative understanding on the effect of Mg by comparing the results of the fit and the In content. Note that in this case, the Al content was set to 0.55 and the In content was then used as a parameter to adjust the fit. The In contents inside the (In,Ga)N layers yielded by this method are reported in Table I. The simulation revealed that, though binary InN was supplied, only (In,Ga)N QWs with thickness of about 1 ML formed. In addition, the maximum In content determined in the (In,Ga)N QW was 0.18, that was obtained for the undoped sample (sample

A). For the samples for which Mg was supplied, the simulation revealed a decrease in the In incorporation. The decrease was the most severe when Mg was supplied simultaneously to In (samples B and E), and the In content inside the (In,Ga)N ML fell below $x = 0.08$. In contrast, for the samples where Mg was supplied in the GaN layers, the In content was found higher with $x = 0.15$ (samples C and D).

Table I: In content in samples A – E determined by the XRD simulation utilizing a fictive (Al,In,Ga)N ML instead of (In,Ga)N:Mg with Al content 0.55 and Nominal Mg concentration derived from SIMS calibration sample. The samples are ordered with increasing In amount from the bottom to the top.

|  | In content in $In_xGa_{1-x}N$ ($x$) | Nominal [Mg] (cm$^{-3}$) |
|---|---|---|
| Sample A | 0.18 | 0 |
| Sample B | 0.05 | $11.8 \times 10^{20}$ |
| Sample C | 0.15 | $7.8 \times 10^{20}$ |
| Sample D | 0.15 | $1.6 \times 10^{20}$ |
| Sample E | 0.08 | $4.1 \times 10^{20}$ |

C. **Secondary ion mass spectroscopy measurements**

Figure 4 illustrates the SIMS measurements of the In and Mg concentration profiles along the first 120 nm from the surface for the SL samples C and E. The In and Mg concentration profiles remarkably exhibited clear oscillations with a period corresponding nearly to the one of the SL. Note that the In and Mg extrema extended over thicknesses much larger than a single ML. This effect is most probably to attribute to the measurement technique that is not suitable to get local concentrations in such a small thickness. For sample C, the oscillations of the In and Mg concentration maxima were out of phase while for sample E they were almost in phase.

For both samples, the In concentration maxima of the oscillations were approximately located at the depth expected for the (In,Ga)N QWs (Figure 4). For sample C, the In concentration reached a maximum level of $2.7 \times 10^{20}$ cm$^{-3}$ in the first QW, then it gradually decreased, and increased again back to a similar level toward the last three QWs. For samples E, the oscillations were very regular and peaked at $9.3 \times 10^{19}$ cm$^{-3}$ except for the last period ($3.8 \times 10^{19}$ cm$^{-3}$) possibly due to surface charging effect.

The In content in the (In,Ga)N layers was calculated by dividing the In concentration integrated along the 10 SL periods and divided by the thickness of the 10 (In,Ga)N ML QWs by the atom site density in GaN ($4.45 \times 10^{22}$ cm$^{-3}$). The results are reported in Table II. For both samples, the In content determined from the SIMS measurement were similar to the one determined by the XRD presented in Table I(a), which supports the validity of the method utilized for the XRD fits.

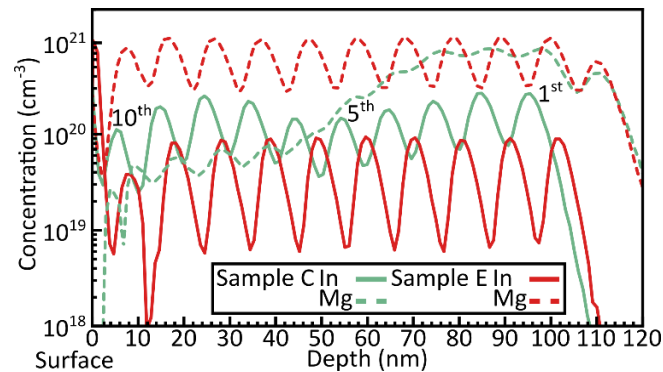

**Figure 4:** Mg and In concentration profiles measured by SIMS for samples C and E. The 1$^{st}$, 5$^{th}$, and 10$^{th}$ SL periods are labeled in the graphs. The abrupt drop near the surface is an artifact of the measurement.

Considering the Mg concentration profiles, both samples presented an additional peak at the interface between the GaN buffer and the first GaN barrier. This peak was located at a depth of around 110 nm from the surface and it most probably resulted from the unintentional incorporation of Mg either accumulated on the manipulator during the growth interrupt necessary to set the growth temperature for the (In,Ga)N and/or flushed back on the sample surface by the N plasma.

Table II: Mg concentration averaged over the whole structure and in the doped segments and In concentration in the (In,Ga)N QWs (samples C, and E) determined by SIMS.

|  | Average Mg concentration (cm$^{-3}$) | Local Mg concentration in the doped segments (cm$^{-3}$) | Mg content in the doped segments - Mg$_y$In$_x$Ga$_{1-x-y}$N ($y$) | In content in the In$_x$Ga$_{1-x}$N ML ($x$) |
|---|---|---|---|---|
| Sample C | $3.2 \times 10^{20}$ | $3.3 \times 10^{20}$ | 0.01 | 0.11 |
| Sample E | $6.3 \times 10^{20}$ | $2.6 \times 10^{22}$ | 0.58 | 0.04 |

For sample C, the Mg concentration oscillated with a concentration of $7.0 \times 10^{20}$ cm$^{-3}$ averaged over the first three periods with a maximum value of $7.7 \times 10^{20}$ cm$^{-3}$. After the third period, the Mg concentration gradually decreased until the seventh period. For the last three periods, the Mg concentration decreased further with a relatively smaller slope down to the value of $4.6 \times 10^{19}$ cm$^{-3}$. This may be due to the fact that Mg accumulated on the surface over the periods inducing a decrease of the Mg sticking coefficient.[34] However, the polarity of the sample might also have changed from Ga-polar to N-polar as the Mg concentration was one order of magnitude higher than the threshold for polarity change reported in Ref. [35]. Note that for this sample, the gradual decrease in the Mg concentration for the barriers closer to the surface was accompanied by a slight increase in the In concentration. In contrast, the Mg concentration in sample E presented regular oscillations which peaked at $1.0 \times 10^{21}$ cm$^{-3}$ and with an amplitude much bigger than for sample C.

The experimental Mg concentration averaged over the whole SL was calculated by integrating the signal along the 10 SL periods and dividing it by the thickness of the SL. The results are presented in Table II. We have also calculated the nominal average Mg concentration for each sample (Table I) based on (In,Ga)N:Mg, or GaN:Mg calibration samples. Note that the nominal average Mg concentration increased with the increasing Mg supply time of the SL doped segment as this calculation disregarded more complex growth mechanisms such as surfactant effect.

For sample C, the average Mg concentration determined by SIMS was two times lower than the nominal Mg concentration reported in Table I. This result may be attributed to the fact that the Mg concentration measured by SIMS drastically changed from the first period to the last period. In comparison, sample E exhibited a Mg concentration that was 1.5 times higher than the nominal one (Table I), which was in fair agreement with our calibration results.

In order to assess the local Mg concentration in the doped SL segments, we have multiplied the average Mg concentration by the thickness of one period and divided it by the thickness of the doped segment. Further, in order to get an estimate of the Mg content in the doped segment, we have divided the Mg concentration in the segment by the atom density in GaN ($4.45 \times 10^{22}$ cm$^{-3}$). The Mg content calculated for sample E was 0.58. This result may be due to the fact that the presence of In increased drastically the incorporation of Mg.[30,36] Note that this increase was accompanied by a significant decrease in the In content (Table II) as also observed by XRD (Table I) and suggested by the QMS measurements. Additionally, the high Mg content of this sample could possibly explain the higher relative intensity of SL$_{+n}$/SL$_{-n}$ revealed by XRD.

## V.   DISCUSSION

Our results clearly evidenced mutual effects of Mg and In for their incorporation in GaN. Figure 5 summarizes schematically the mechanisms at play during the supply of In and Mg in presence of N as inferred by QMS.

When solely In and N are supplied (experiments 4 and A – case 1 in Figure 5) on GaN(0001), In atoms chemisorb and the $^{115}$In desorption signal rises gently at the opening of the In cell. With continued In supply, In adatoms reach a maximum surface coverage limited by the substrate temperature[37,38] and the $^{115}$In desorption signal reaches a maximum level. With continued supply, In accumulates in a liquid metallic phase and the weakly bonded excess of In segregates on the growth front and desorbs during the subsequent deposition of GaN or/and the following growth interruption, as already explained in Refs. [19] and [39]. Note that this mechanism is unchanged when In and N are supplied on Mg-doped GaN(0001) (Experiments 6, C and D) unless Mg already accumulated at the surface before supplying InN.

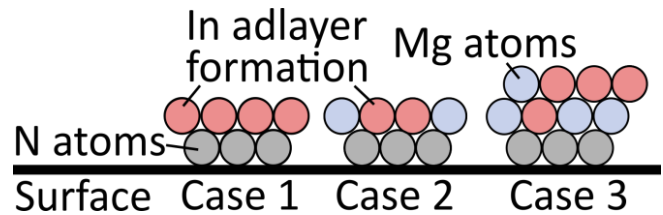

**Figure 5:** Atomic model of the "In supply" stage for the 3 different cases described in the discussion. For case 1, during the In supply, impinging In atoms form either an In adlayer or desorb from the surface. For case 2, the simultaneous supply of Mg reduces the number of free N bonds available for In, and prevents the formation of an In adlayer. For case 3, Mg adatoms accumulated on the surface modify the In coverage. The In, N and Mg atoms are highlighted with red, gray, and blue, respectively.

When In, Mg and N are simultaneously supplied on GaN(0001) (experiments 5, first period of B, and E – case 2 in Figure 5), because Mg and In occupy the same cation site in GaN, these two elements compete for the limited free surface sites. However, the incorporation of Mg is kinetically favored compared to In for two reasons: First, In is a surfactant for Mg as it decreases the binding energy of Mg in GaN, and so it increases Mg incorporation in GaN.[5,27,30] Second, as Mg accumulates at the surface, it possibly inhibits the formation of an In adlayer, similarly to Ga[32] and it promotes In desorption from the surface. Consequently, the incorporation of Mg in GaN and, as evidenced by QMS, the desorption of In from the surface increase leading to a decrease of the In incorporation confirmed by XRD and SIMS. Note that the reduction of the In incorporation in thick (In,Ga)N layers in presence of Mg was also reported by Gherasoiu *et al.*[40] and attributed to a kick-out mechanism due to the reduction of the number of excess electrons by substitutional Mg in the In or Ga-terminated surfaces. Besides, an increase of Mg incorporation in AlN by Mg and In codoping was also reported and attributed to the ionization of Mg atoms by In-$V_N$ complexes.[36] We speculate that a similar electronic interaction could also enhance the incorporation of Mg into (In,Ga)N.

In the last case when In is supplied on a GaN surface where a Mg adlayer already accumulated (experiments 3, possibly B from the second period on, C and D – case 3 in Figure 5), more In can be accommodated by the Mg adlayer, possibly forming a liquid metallic phase with increasing In content at 610 °C[41] and therefore the In coverage is modified. Note that these In atoms are only weakly bonded and desorb concurrently.

## VI. CONCLUSIONS

We have investigated the influence of Mg on the In adsorption and incorporation in (In,Ga)N superlattices grown on GaN(0001) by PAMBE at 610°C. *In-situ* measurements by QMS revealed that the In desorption decreased in the presence of Mg only. In contrast, it increased in the presence of N and Mg. Furthermore, different (In,Ga)N/GaN SL samples were grown to study these effects by XRD and SIMS. Both, the *in-situ* and *ex-situ* results indicated the decrease of In incorporation during the simultaneous supply of In and Mg under N supply.

XRD scans around GaN(0002) revealed that the maximum In content estimated by the simulation of the spectra was 0.18 in the undoped SL sample and it decreased down to 0.05 when Mg was supplied simultaneously to In. Additionally, the compressive stress of Mg in the SL GaN layers was revealed by the increased intensity of the $SL_{+n}$ satellite peaks.

A clear increase of the Mg concentration in the presence of In was evidenced by SIMS measurements of the SLs and a maximum Mg content of 0.58 was determined if it was to incorporate in a single (In,Ga)N ML. Furthermore, this measurement confirmed that the In content decreased markedly for the all the samples in which Mg was supplied simultaneously to In. These results confirmed the increase of the In desorption observed by QMS and the decrease of the In content in the (In,Ga)N/GaN SL measured by XRD.

These findings are direct results of the competition between Mg and In for their incorporation at the same cation sites on the surface of GaN(0001). As In acts as a surfactant for Mg incorporation in GaN, it leads to an increase of the Mg content but the decrease of the In content in (In,Ga)N/GaN SLs. However, when a metallic Mg adlayer forms, it may actually accommodate more In that is only weakly bond and desorbs concurrently.

These results reveal that it is essential to modulate the In and Mg supplies in order to form (In,Ga)N SLs with high In and Mg content. This understanding might improve the efficiency of the p-type doping of (In,Ga)N SLs by the design of heterostructures with optimum In and Mg content for higher hole injection efficiency.

## ACKNOWLEDGMENTS

The authors thank H. P. Schönherr for the maintenance of the MBE chamber, RTG Mikroanalyse GmbH for SIMS characterizations, and T. Auzelle for the critical review of the manuscript. Funding of this work by European Union's Horizon 2020 research and innovation program (Marie Skłodowska-Curie Actions) under Grant Agreement "SPRING" No. 642574 is gratefully acknowledged.

## DATA AVAILABILITY

The data that supports the findings of this study are available within the article.